\documentclass{revtex4}

\usepackage[latin1]{inputenc}
\usepackage{graphicx}
\usepackage{color}
\usepackage{psfrag}
\usepackage{amssymb}

\newcommand{\eq}[1]{\begin{equation} #1 \end{equation}}
\newcommand{\eqa}[1]{\begin{eqnarray} #1 \end{eqnarray}}
\newcommand{\sss}{\scriptscriptstyle}

\def\lg{^{\rm long}}
\def\Amix{\mathcal{A}_{\rm mix}}
\def\Adir{\mathcal{A}_{\rm dir}}

\begin{document}

\date{\today}

\preprint{LPT-ORSAY/07-}
\preprint{yyyy}

\title{Penguin-mediated $B_{d,s}\to VV$ decays\\ and the $B_s-\bar{B}_s$
  mixing angle}

\author{S\'ebastien Descotes-Genon$^{a}$, Joaquim Matias$^{b}$ and Javier Virto$^{b}$}
\affiliation{
$^{a}$ Laboratoire de Physique Th\'eorique, CNRS/Univ. Paris-Sud 11 (UMR 8627), 91405 Orsay Cedex, France\\
$^{b}$ IFAE, Universitat Aut\`onoma de Barcelona, 08193 Bellaterra, Barcelona, Spain}

\begin{abstract}
In this letter, we propose three different strategies to extract
the weak mixing angle $\phi_s$ of the $B_s$ system using
penguin-mediated decays into vectors, mainly $B_s \to K^{*0}
{\bar K}^{*0}$, $B_s \to \phi{\bar K}^{*0}$ and $B_s \to
\phi\phi$. We also provide predictions for the longitudinal
branching ratio and CP-asymmetries of $B_s \to K^{*0} {\bar
K}^{*0}$ using a method that combines QCD factorisation with
flavour symmetries to relate this decay to its $B_d$ counterpart.

\end{abstract}

\pacs{???}

\maketitle

\section{INTRODUCTION} \label{sec:intro}

The large amount of data collected by Babar and Belle, the
progress of CDF and D0 and the advent of LHCb have increased our
ability of testing the CP- and flavour-violating
structure of the Standard Model (SM), increasing our chances of
discovering New Physics. The phenomenology of penguin-dominated
hadronic $B$-decays is particularly relevant in this field. At the
theoretical level, $b \to s$ penguin transitions are expected to
receive a large impact from New Physics~\cite{nir}, compared to $b
\to d$ penguin transitions. This comparison has driven detailed
experimental and theoretical analyses devoted to $B$-decays that
proceed through a $b\to s$ transition, such as
 $B\to \pi K$ \cite{BpiKex,BpiKth}, $B\to(\phi,\eta,\omega,\rho\dots)K_s$
\cite{BVKex,BVKth}\ldots Deviations from theoretical expectations
were observed for some of these decays, but it remains unclear
whether these results can or cannot be explained within the SM
\cite{probs1,probs2}.

The properties of the $B_s$ meson have attracted a lot of
attention recently : the $B_s-\bar{B}_s$ mass difference $\Delta
M_s$ has been measured~\cite{DeltaMs},  with an immediate impact
on New Physics~\cite{thDeltaMs}. Very recently, first
experimental information on $\Delta \Gamma$ and the ${\rm arg}(-M_{12}/\Gamma_{12})$
of
the
$B_s$ system has also been presented~\cite{expphase} (see also \cite{theory}). Another important
piece of information will come with the measurement of the
mixing
angle $\phi_s$. Its SM value is  $\phi_s^{SM}=2\beta_s=-2\lambda^2
R_b\sin{\gamma}\simeq -2^\circ$~\cite{Buras}
and thus probes new CP violating phases with a high sensitivity.
As far as decay modes are concerned, a considerable amount of
theoretical work has been carried out to understand charmless
$B_s$ decays (for instance in $B_s \to KK$ decays \cite{lar,fm}).
The experimental program,  focusing initially on $B_s\to\pi K$ and
$B_s\to KK$ modes~\cite{Bsexp}, will extend its scope by
considering more and more classes of $B_s$ decays. More
observables could be studied on such details
with the recent prospect of $B$ or super-$B$
factories working at the $\Upsilon(5S)$ and thus producing $B_s$-$\bar{B}_s$ pairs~\cite{Belle5S}.

On the theoretical side, non-leptonic $B_s$ decays have been extensively studied,
with a recent emphasis on the case of two final vector mesons ($B_s\to VV$), within the
framework of QCD factorisation (QCDF)~\cite{benekeBVV} (see ref.~\cite{ali}
for related work within the pQCD approach) and in the
context of $SU(3)$ flavour symmetries~\cite{silvestrini}. QCDF and flavour
symmetries provide different tools to tackle non-leptonic decays, with their
advantages and shortcomings : the former is a systematic expansion in $1/m_b$ but encounters
difficulties with phenomenology due to power-suppressed hadronic effects, such as
 final-state interactions.
The latter takes hadronic effects into account but may be affected
by large corrections, up to 30 \% for $SU(3)$ relations.

In ref.~\cite{DMV} (see also \cite{BLMV,DILM,Virto:2007cb}), we have developed
an intermediate approach that aims at combining $SU(3)$ relations
with QCDF-inspired input in a theoretically controlled way, in
order to gain precision over the other approaches. In this letter
we extend this approach to a larger class of decay modes and apply
it mainly to $B_s\to K^{*0}\bar{K}^{*0}$ and partially to
$B_s\to \phi \bar{K}^{*0}$ and $B_s \to\phi\phi$ decays. These
decays exhibit particularly alluring experimental and theoretical
features. For instance, on the experimental side,
$B_s\to K^{*0}\bar{K}^{*0}$
is likely to be measured easily in hadronic machines, due to its
expected high branching ratio $\sim\mathcal{O}(10^{-5})$
and its prospects are being analyzed in detail at present at LHCb. In
addition, some decay products of the $K^*$ resonances are charged
($\pi^{\pm},K^{\mp}$) and are easier to identify experimentally than
in the case of $B_s\to K^0\bar{K}^0$. On the
theoretical side, these decays allow for an accurate extraction of
the $B_s-\bar{B}_s$ mixing angle, with a small direct CP asymmetry in the SM
that should be very sensitive to CP-violating New Physics.

The aim of this letter is to provide three strategies to extract $\phi_s$ from
certain non-leptonic B decays. It is organised as follows. In Section
\ref{sec:longobs}, we consider longitudinal observables for $B\to VV$ decays, and
relate them to the observables usually obtained from an angular
analysis of such decays. In Section \ref{sec:pt}, we describe a
general method to extract the SM hadronic parameters of a $B$
meson decay given the branching ratio and a theoretical quantity
called $\Delta$, which is the difference between tree and penguin
contributions (see \cite{DMV}). We also derive useful bounds for
the branching ratios and direct CP asymmetries as a function of
$\Delta$. Section \ref{sec:Delta} deals with the theoretical
computation of $\Delta$ for $B_{d,s}\to K^{*0} \bar{K}^{*0},\phi
\bar{K}^{*0},\phi\phi$ decay modes : for penguin-dominated modes,
this difference is dominated by short distances and can be
computed accurately in QCD factorisation. In Section
\ref{sec:bounding}, we exploit this theoretical information to put
a bound on the tree pollution affecting the determination of CKM
angles through the mixed CP-asymmetry. In Section
\ref{sec:measuring}, we present general expressions to extract the
CKM phases $\alpha$, $\beta$, $\gamma$ and $\beta_s$ from hadronic
penguin-dominated $B_{d,s}$ decays. These general expressions can
be applied to $B_{d,s}\to K^{*0}\bar{K}^{*0}$, $B_{d,s}\to
\phi \bar{K}^{*0}$ and $B^0_{s}\to \phi \phi$ decay modes. In
Section \ref{sec:flavour}, we use flavour symmetries and QCD
factorisation to relate $B_d\to K^{*0} \bar{K}^{*0}$ and $B_s\to
K^{*0} \bar{K}^{*0}$ observables, and exploit this $U$-spin
symmetry to constrain the $B_s-\bar{B}_s$ mixing. Finally, we
discuss the three strategies developed and we conclude in Section~\ref{sec:discussion}.

\section{LONGITUDINAL OBSERVABLES IN $B\to VV$ MODES} \label{sec:longobs}

The amplitude for a $B$ meson decaying into 2 vector mesons can be written as
\eqa{
A(B\to V_1V_2)&=&\left[ \frac{4m_1m_2}{m_B^4}(\epsilon_1^*\cdot p_B)(\epsilon_2^*\cdot p_B) \right] A_0 \nonumber\\
&+&\left[\frac{1}{2}(\epsilon_1^*\cdot\epsilon_2^*)
-\frac{(p_B\cdot\epsilon_1^*)(p_B\cdot\epsilon_2^*)}{m_B^2}
-\frac{i\epsilon_{\mu\nu\rho\sigma}\epsilon_1^{*\mu}\epsilon_2^{*\nu}p_1^\rho p_2^\sigma}{2 p_1\cdot p_2}\right] A_+\nonumber\\
&+&\left[\frac{1}{2}(\epsilon_1^*\cdot\epsilon_2^*)
-\frac{(p_B\cdot\epsilon_1^*)(p_B\cdot\epsilon_2^*)}{m_B^2}
+\frac{i\epsilon_{\mu\nu\rho\sigma}\epsilon_1^{*\mu}\epsilon_2^{*\nu}p_1^\rho p_2^\sigma}{2 p_1\cdot p_2}\right] A_-
}
where $A_{0,+,-}$ correspond to the amplitudes for
longitudinal and transversely polarized final vector mesons.
It is also customary to use the basis $A_{0,\|,\bot}$, where
$A_{\|,\bot}=(A_+\pm A_-)\sqrt{2}$.

The vector mesons in the
final state decay typically into pairs of pseudoscalar particles.
A full angular analysis of vector-vector modes provides the
following set of observables: three polarisation fractions $f_0$,
$f_\bot$ and $f_\|$ (only two of them are independent) and their
CP-conjugate counterparts $\bar{f}_{0,\bot,\|}$, two phases
$\phi_{\bot,\|}$ (again, together with $\bar{\phi}_{\bot,\|}$), a
total CP-averaged branching ratio $BR$, and a total direct CP-asymmetry
$\Adir$. The polarisation fractions are defined as
\eq{
f_{0,\bot,\|}\equiv\frac{|A_{0,\bot,\|}|^2}{|A_0|^2+|A_\bot|^2+|A_\||^2}  \quad \quad
{\bar f}_{0,\bot,\|}\equiv\frac{|\bar{A}_{0,\bot,\|}|^2}{|\bar{A_0}|^2+|\bar{A_\bot}|^2+|\bar{A_\|}|^2}
}
A full angular analysis is available for $B_d\to\phi K^{*0}$
from BaBar and Belle \cite{babar phiK,belle phiK},
and the same type of analysis is expected for $B_d\to K^{*0} \bar{K}^{*0}$.

We will focus in this paper on observables for the longitudinal
polarisation ($BR\lg$, $\Adir\lg$, $\Amix\lg$ and
$\mathcal{A}_{\Delta\Gamma}\lg$), where only $A_0$ occurs. These
observables, free from the positive and negative helicity
components, can be predicted with a much better accuracy. Indeed
the negative-helicity (positive-helicity) component of the
amplitude is $1/m_b$-suppressed ($1/m_b^2$-suppressed) because of
the nature of the interactions involved (left-handed weak
interaction, helicity-conserving strong interaction at high
energies)~\cite{KaganBVV,benekeBVV}. This suppression makes
longitudinal observables better behaved and easier to compute than
transverse ones.

Some decay channels exhibit the $1/m_b$-suppression of transverse amplitudes
in a very striking way : the longitudinal polarisation is very close to 1, e.g. $f_L\simeq 97\%$ for
$B\to \rho^+\rho^-$. In such cases, the full observables (where
$A_0$ is replaced by the sum $A=A_0+A_-+A_+$) coincide with the longitudinal
ones to a high degree of accuracy. On the other hand,  for
penguin dominated $\Delta S=1$ decays, $f_L$ can be as low as $\sim 50\%$,
so that the transverse amplitudes (or $\pm$ helicity amplitudes)
contribute significantly to the full observables. Therefore, one must
determine whether purely longitudinal observables can be extracted from
experimental measurements.

We start from the normalized partial decay rate of $B\to V_1V_2$,
where the two vector mesons go subsequently into pairs of
pseudoscalar mesons. It can be written~\cite{BtoVVcrosssect}
\begin{eqnarray}
\frac{d^3\Gamma}{\Gamma d\cos\theta_1 d\cos\theta_2 d\phi}&=&
 \frac{9}{8\pi}\frac{1}{|A_0|^2+|A_\| |^2+|A_\bot |^2}\\ \nonumber
&&
 \times
  \Bigg[|A_0|^2 \cos^2\theta_1 \cos^2\theta_2
       +|A_\| |^2 \frac{1}{2}\sin^2\theta_1 \sin^2\theta_2 \cos^2\phi \\ \nonumber
&& \    +|A_\bot |^2 \frac{1}{2}\sin^2\theta_1 \sin^2\theta_2 \sin^2\phi
       +{\rm Re}[A_0^* A_\| ] \frac{1}{2\sqrt{2}}\sin 2\theta_1 \sin 2\theta_2 \cos\phi \\ \nonumber
&& \
       +{\rm Im}[A_0^* A_\bot ] \frac{-1}{2\sqrt{2}}\sin 2\theta_1 \sin 2\theta_2 \sin\phi
       +{\rm Im}[A_\|^* A_\bot ] \frac{-1}{2}\sin^2\theta_1 \sin^2\theta_2\sin 2\phi
  \Bigg]
\end{eqnarray}
where $(\theta_1,\theta_2,\phi)$ are angles introduced to
describe the kinematics of the decay $B\to V_1 V_2$ followed by
$V_1 \to P_1 P'_1$ and $V_2 \to P_2 P'_2$. $\theta_1$ is the
angle of one of the $V_1$ decay products in the rest frame of $V_1$
relative to the motion of $V_1$ in the rest frame of the $B$-meson (same
for $\theta_2$ with $V_2$). $\phi$ is the angle between the two planes formed
by the decay products of $V_1$ and $V_2$ respectively (see for instance Fig.1 of ref.~\cite{BtoJPsiKstar} for
a representation of the angles).

There are different ways to perform the angular integrations in order to
extract the purely longitudinal component from the differential decay rate.
A first option consists in computing moments of $\cos\theta_1$ (or
equivalently $\cos\theta_2$) :
\begin{equation}
\Gamma\lg
  \equiv \int \frac{d^3\Gamma}{d\cos\theta_1 d\cos\theta_2 d\phi}
      \left(\frac{5}{2}\cos^2\theta_1-\frac{1}{2}\right)
    d\cos\theta_1 d\cos\theta_2 d\phi = g_{PS} |A_0|^2/\tau_B
\end{equation}
where $g_{PS}$ is the product of phase-space and lifetime factors
\begin{equation} \label{gps}
g_{PS}=\frac{\tau_B}{16\pi
M_B^3}\sqrt{[M_B^2-(m_1+m_2)^2][M_B^2-(m_1-m_2)^2]}
\end{equation}
A second possibility amounts to performing asymmetric integrations over one angle~\cite{sharma}
\begin{equation}
\Gamma\lg
  \equiv
 \int_{-1}^{1}  d\cos\theta_1 \int_T d\cos\theta_2
 \int_{0}^{2\pi} d\phi
 \frac{d^3\Gamma}{d\cos\theta_1 d\cos\theta_2 d\phi}
  = g_{PS} |A_0|^2/\tau_B
\end{equation}
with
\begin{equation}
\int_T d\cos\theta_2=\left(\frac{11}{9}\int_0^{\pi/3}-\frac{5}{9}\int_{\pi/3}^{2
\pi/3}+\frac{11}{9}\int_{2 \pi/3}^{\pi} \right) (-\sin\theta_2) d\theta_2
\end{equation}
In the same way we can obtain the CP-conjugate $\Gamma\lg({\bar
B_q^0} \to \bar{f})$ from the corresponding CP-conjugate distribution, leading
to the CP-averaged branching ratio of the longitudinal component
\begin{equation}
BR\lg =
 \frac{\tau_B}{2} \left(
 {\Gamma\lg(B^0_q\to f)
              +\Gamma\lg (\bar{B}^0_q\to \bar{f})} \right)
 = g_{PS} \frac{|A_0|^2+|\bar{A}_0|^2}{2}
\label{BR}
\end{equation}
where $\bar{A}_0$ is the CP-conjugate amplitude of $A_0$.

If we include the dependence on time in the above expressions, $B$-$\bar{B}$ mixing modifies
the expressions~\cite{BtoJPsiKstar}. We will focus on CP-eigenstates $f_{\rm \sss CP}$ in the final state
$K^{*0} \bar{K}^{*0}$ and $\phi\phi$, as well as $\phi K^{*0}$ with a subsequent
decay of
$K^{*0}$ into a CP-eigenstate ($K_s \pi^0$ or $K_L \pi^0$).

The time evolution of these observables is obtained by considering the
time dependence of $A_0(t)$~\cite{chiang}. Inserting this
time dependence one arrives at the usual expression for the
longitudinal component of the time-dependent CP-asymmetry:
\eq{ \mathcal{A}_{\rm CP}(t) \equiv \frac{\Gamma\lg (B^0_q(t)\to
f_{\rm \sss CP})-\Gamma\lg (\bar{B}^0_q(t)\to f_{\rm \sss CP})}{\Gamma\lg (B^0_q(t)\to
f_{\rm \sss CP})+\Gamma\lg (\bar{B}^0_q(t)\to f_{\rm \sss CP})}
=\frac{\Adir\lg\cos{(\Delta Mt)}+\Amix\lg\sin{(\Delta Mt)}}
{\cosh{(\Delta\Gamma
t/2)}-\mathcal{A}\lg_{\Delta\Gamma}\sinh{(\Delta\Gamma t/2)}} }
where the direct and mixing-induced CP asymmetries
are defined by:
\eq{ \Adir\lg\equiv
\frac{|A_0|^2-|\bar{A}_0|^2}{|A_0|^2+|\bar{A}_0|^2},\quad
\Amix\lg\equiv -2 \eta_f \frac{{\rm
Im}(e^{-i\phi_M}A_0^{*}\bar{A_0})}{|A_0|^2+|\bar{A_0}|^2}
\label{CPAs} }
together with the asymmetry related to the width difference :
\eq{ \mathcal{A}\lg_{\Delta\Gamma}\equiv -2 \eta_f\frac{{\rm
Re}(e^{-i\phi_M}A_0^{*}\bar{A_0})}{|A_0|^2+|\bar{A_0}|^2}
\label{CPADelta} }
$\phi_M$ is the mixing angle and $\Delta
\Gamma=\Gamma^H-\Gamma^L$. $\eta_f$ is the CP eigenvalue of the
final state $f$ ($\pm 1$): $\eta_{K^{*0}K^{*0}}=\eta_{\phi\phi}=1$,
whereas $\eta_{K^{*0}
  \phi}=1$ if $K^{*0}$ decays into $K_s\pi^0$ and $-1$ if it decays into
  $K_L\pi^0$. In the latter case, the contribution from the strong process
$K^{*0}\to K\pi$ is the same for both $B$ and $\bar{B}$ decays
and it cancels in the time-dependent CP-asymmetry Eq.(\ref{CPAs}),
which depends only on the amplitudes $A_0$ and $\bar{A}_0$.



Finally, if the direct CP-asymmetries of all three
helicity components are
negligible, the longitudinal branching ratio can be estimated very easily from:
$BR\lg=BR^{\rm total} f_0$.


\section{DETERMINING PENGUIN AND TREE CONTRIBUTIONS} \label{sec:pt}

We consider a $B$ meson decaying through
$\bar{b}\to\bar{D}q\bar{q}$, with $D={d,s}$, and restrict the
discussion to the longitudinal component of the amplitude.
However, the results are general and can be applied to any $B \to
PP, PV, VV$ decay (in the latter case, the relations hold for each helicity
amplitude independently). We can parameterize the
amplitudes in
terms of ``tree'' and ``penguin'' contributions,
\eq{
A_0=\lambda_u^{(D)*}T + \lambda_c^{(D)*}P\ ,\quad \bar{A}_0=\lambda_u^{(D)}T + \lambda_c^{(D)}P.
\label{As}
}
where $\lambda_U^{(D)}\equiv V_{UD}^*V_{Ub}$ are combination of CKM
elements, $U={u,c}$. The penguin and tree contributions are defined
through their associated CKM factor, and not from the topology of
the relevant diagrams (even though in many cases, tree contributions
correspond to tree diagrams).
Such a decomposition is always possible and completely general in the
Standard Model since the unitarity of the CKM matrix allows to recast
contributions proportional to $V_{tD}^*V_{tb}$ into the form of eq.~(\ref{As}). We
will follow the convention of calling ``penguin'' the piece proportional
to $V_{cD}^*V_{cb}$ and ``tree'' the piece proportional to $V_{uD}^*V_{ub}$. In the particular case
of penguin-mediated decays, there is no actual tree diagram and the
tree contribution corresponds to penguins containing a $u$-quark loop
(or a $t$-quark loop).

For both neutral and charged decays, one can define the
CP-averaged branching ratio and the direct CP asymmetry as given in
(\ref{BR})
 and (\ref{CPAs}). From these two observables we can obtain the magnitudes of the
amplitudes
\begin{equation}
|A_0|^2 = BR\lg(1+\Adir\lg)/g_{PS} \qquad |\bar{A_0}|^2 =
BR\lg(1-\Adir\lg)/g_{PS} \label{eqsBRAdir}\end{equation}

We consider the quantity $\Delta$ defined as the difference between tree and penguin
hadronic contributions~\cite{DMV}
\eq{\label{delta}\Delta\equiv T-P}
The value of $\Delta$ might be determined on theoretical grounds, for instance
 through QCD factorisation~\cite{BBNS}. In the next sections, we will consider
decays where such a computation is particularly clean and free from many
 long-distance uncertainties.
Given the arbitrary common phase for $T$ and $P$, we can always rotate
simultaneously $P$ and $\Delta$ and choose $\Delta$ to be real positive if we restrict
ourselves to a given channel. We will adopt this
convention in the following, unless the contrary is explicitly stated.

We can write down the amplitudes (\ref{As}) in the following way:
\eq{
\begin{array}{rcl}
|A_0|^2 & = & |\lambda_c^{(D)*}+\lambda_u^{(D)*}|^2\left| P +
\frac{\lambda_u^{(D)*}}{\lambda_c^{(D)*}+\lambda_u^{(D)*}} \Delta \right|^2\\
&&\\
|\bar{A}_0|^2 & = & |\lambda_c^{(D)}+\lambda_u^{(D)}|^2\left| P +
\frac{\lambda_u^{(D)}}{\lambda_c^{(D)}+\lambda_u^{(D)}} \Delta \right|^2
\end{array}
\label{eqsP}
}
The previous equations can be solved for $P$ if we know $BR\lg$,
$\Adir\lg$, $\Delta$, and the CKM parameters $\lambda_u^{(D)}$ and
$\lambda_c^{(D)}$. The solutions exhibit a very simple form for
$\Delta$ real and positive
\eqa{
{\rm Re}[P] & = & -c_1^{(D)}\,\Delta \pm \sqrt{-{\rm
Im}[P]^2-\left(\frac{c_0^{(D)}\Delta}{c_2^{(D)}}\right)^2+\frac{\widetilde{BR}}{c_2^{(D)}}}
\label{eqxP}\\
{\rm Im}[P] & = & \frac{\widetilde{BR}\,\Adir\lg}{2
  c_0^{(D)}\Delta}\label{eqyP}
}
where the coefficients $c_i^{(D)}$ are given by
\eq{c_0^{(D)}=\lambda_c^{(D)}|\lambda_u^{(D)}|\sin{\gamma}\,;\ \
c_1^{(D)}=(|\lambda_u^{(D)}|^2+\lambda_c^{(D)}|\lambda_u^{(D)}|\cos{\gamma})/c_2^{(D)}\,;\ \
c_2^{(D)}=|\lambda_u^{(D)}+\lambda_c^{(D)}|^2}
and $\widetilde{BR}\equiv BR\lg/g_{PS}$. The numerical values of
these coefficients are collected in Table~\ref{coeffs}. Once $P$
is known, Eqs.~(\ref{delta}), (\ref{eqyP}) and (\ref{eqxP}) yield
the second hadronic parameter $T$.
\begin{table}
\begin{tabular}{cccccccc}
\hline
$c_0^{(d)}$ & $c_1^{(d)}$ & $c_2^{(d)}$ & $c_0^{(s)}$    &    $c_1^{(s)}$   &   $c_2^{(s)}$  &  $\mathcal{R}_d$
  &  $\mathcal{R}_s$\\
\hline
$\ -3.15\cdot 10^{-5}\ $  &  $\ -0.034\ $  &  $\ 6.93\cdot 10^{-5}\ $  &  $\ 3.11\cdot 10^{-5}\ $  &  $\ 0.011\ $
 &  $\ 1.63\cdot 10^{-3}\ $
& $7.58\cdot 10^{-3}$  &  $1.54\cdot 10^{-3}$ \\
\hline
\end{tabular}
\caption{Numerical values for the coefficients $c_i^{(D)}$ and $\mathcal{R}_D$ for $\gamma=62^\circ$.}
\label{coeffs}
\end{table}

Interestingly, two consistency conditions exist between $BR\lg$,
$\Adir\lg$ and $\Delta$, to guarantee the existence of solutions
for $P$ (the argument of the square root in ${\rm Re}[P]$ must be
positive):
\eq{
\begin{array}{rcl}
\displaystyle |\Adir\lg| & \le & \displaystyle \sqrt{\frac{\mathcal{R}_D^2\Delta^2}
{2\widetilde{BR}}\Big(
2-\frac{\mathcal{R}_D^2\Delta^2}{2\widetilde{BR}} \Big)}
\ \ \approx \frac{\mathcal{R}_D \Delta}{\sqrt{\widetilde{BR}}}\\
&&\\
\displaystyle \widetilde{BR} & \ge & \displaystyle \frac{\mathcal{R}_D^2\Delta^2}{4}
\end{array}
\label{consist}}
with the combination of CKM factors
$\mathcal{R}_D=2|c_0^{(D)}|/\sqrt{c_2^{(D)}}$. The approximation
for the upper bound on $|\Adir\lg|$ holds up to very small
corrections in the usual situation $\Delta \lesssim
\mathcal{O}(10^{-7})$ and $BR\lg\sim \mathcal{O}(10^{-6})$.

\begin{figure}
\begin{center}
\psfrag{BR}{\hspace{-1.5cm}$BR\lg(B_d\to K^{*0} \bar{K}^{*0})\times 10^6$}
\psfrag{A}{\hspace{-1.5cm}$\Adir\lg(B_d\to K^{*0} \bar{K}^{*0})$}
\psfrag{-0.4}{\hspace{0.15cm}$-0.4$}\psfrag{-0.2}{\hspace{0.15cm}$-0.2$}\psfrag{0.2}{\hspace{0.15cm}$0.2$}\psfrag{0.4}{\hspace{0.15cm}$0.4$}
\psfrag{0}{$0$}\psfrag{2}{$2$}\psfrag{4}{$4$}\psfrag{6}{$6$}\psfrag{8}{$8$}
\includegraphics[width=10cm]{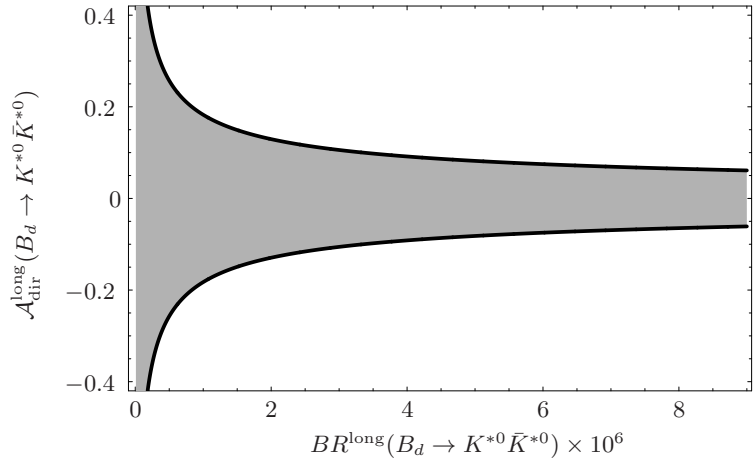}
\end{center}
\caption{Allowed region on the $BR\lg-\Adir\lg$ plane for $B_d\to K^{*0} \bar{K}^{*0}$, according to the value of $\Delta^{d}_{K^*K^*}$.}
\label{BRvsAdirKK}
\end{figure}

The relations derived in this section apply to all charmless
hadronic $B$ meson decays, and are thus of quite generic nature.
As an illustration, we anticipate the results of next section and
assume that we are able to compute $\Delta$ accurately for the
decay $B_d\to K^{*0} \bar{K}^{*0}$ (denoted by
$\Delta^{d}_{K^*K^*}$). Given a measured value for the
longitudinal branching ratio, the quantity $\Delta^{d}_{K^*K^*}$
in Eq.(\ref{delta0KK}) constrains the direct CP asymmetries
according to Eq.(\ref{consist}). The allowed values for the
asymmetry are shown in Fig.~\ref{BRvsAdirKK}.

\section{THE THEORETICAL INPUT OF $\Delta=T-P$} \label{sec:Delta}

The quantity $\Delta$ is a hadronic, process-dependent,
intrinsically non-perturbative object, and thus difficult to
compute theoretically. Such hadronic quantities are usually
extracted from data or computed using some factorization-based
approach. In the latter case, $\Delta$ could suffer from the usual
problems related to the factorisation ansatz and in particular
long-distance effects.

However, for penguin-mediated decays, $T$ and $P$ share the same
long-distance dynamics: the difference comes from the ($u$ or $c$)
quark running in the loops~\cite{DMV}. Indeed, in such decays, $\Delta=T-P$ is not
affected by the breakdown of factorisation that affects
annihilation and hard-spectator contributions, and it can be
computed in a well-controlled way leading to safer predictions and
smaller uncertainties.

For vector-vector final states,
a $\Delta$ is associated to each helicity amplitude, but we focus on
longitudinal quantities here. We obtain for the longitudinal $\Delta$ of the
$B_d\to K^{*0}\bar{K}^{*0}$ ($B_s\to K^{*0}\bar{K}^{*0}$)
decay
denoted  by $\Delta^{d}_{K^*K^*}$ ($\Delta^{s}_{K^*K^*}$):
\eqa{
|\Delta^d_{K^*K^*}|&=&A_{K^*K^*}^{d,0} \frac{C_F \alpha_s}{4\pi N_c}C_1\,|\bar{G}_{K^*}(s_c)-\bar{G}_{K^*}(0)|=(1.85 \pm 0.79)\times 10^{-7}\ {\rm GeV} \label{delta0KK}\\
|\Delta^s_{K^*K^*}|&=&A_{K^*K^*}^{s,0} \frac{C_F \alpha_s}{4\pi N_c}C_1\,|\bar{G}_{K^*}(s_c)-\bar{G}_{K^*}(0)|=(1.62 \pm 0.69)\times 10^{-7}\ {\rm GeV} \label{delta0sKK}
}
where $\bar{G}_V\equiv G_V-r_{\chi}^V \hat{G}_V$ are the usual
penguin functions and $A_{V_1 V_2}^{q,0}$ are the naive
factorization factors combining decay constants and form factors
(see \cite{BN} for definitions),
\eq{A_{V_1 V_2}^{q,0}=\frac{G_F}{\sqrt{2}} m_{B_q}^2 f_{V_2} A_0^{B_q\to V_1}(0)}
The numerical values of the used inputs are given in Table~\ref{inputs}.
The contributions to each error from the various sources are detailed in
Table~\ref{errs}. For the $\Delta$, as well as for the other quantities
computed in this paper, we quote as the central value the value obtained from
taking the central value of the inputs. To estimate the error, we vary one by one each
of the inputs, compute the difference with the central value, then add in
quadrature the resulting uncertainties. The main sources of uncertainties are the scale of
factorisation $\mu$, the mass of the charm quark $m_c$, and the form factor $A_0^{B\to K*}$.

\begin{table}
\begin{tabular}{cccccccc}
\hline
$m_c(m_b)$  &    $f_B$       &    $f_{B_s}$   & $\lambda_B,\lambda_{B_s}$ & $\alpha_1^{(\bot)}(K^*)$  & $\alpha_2^{(\bot)}(K^*)$  &   $f_{K^*}$    \\
\hline\hline
$\ 1.3\pm 0.2\ $ & $\ 0.21\pm 0.02\ $ & $\ 0.24\pm 0.02\ $ &     $\ 0.35\pm 0.15\ $        &     $\ 0.06\pm 0.06\ $        &          $\ 0.1\pm 0.2\ $     & $\ 0.218\pm 0.004\ $\\
\hline
\end{tabular}\\
\vspace{0.3cm}
\begin{tabular}{cccccccc}
\hline
$f_{K^*}^\bot(2{\rm GeV})$ & $A_0^{B\to K^*}$  & $A_0^{B_s\to K^*}$  &  $f_\phi$          &  $f_\phi^\bot(2{\rm GeV})$  &  $A_0^{B_s\to \phi}$       & $\alpha_2^{(\bot)}(\phi)$  \\
\hline\hline
     $\ 0.175\pm 0.025\ $      &  $\ 0.39\pm 0.06\ $   & $\ 0.33\pm 0.05\ $      & $\ 0.221\pm 0.003\ $   &  $\ 0.175\pm 0.025\ $           &  $\ 0.38^{+0.10}_{-0.02}\ $  & $\ 0.0\pm 0.3\ $  \\
\hline
\end{tabular}
\caption{Input parameters required in QCD factorisation to compute the
  quantities $\Delta$'s and $\delta$'s described in the text. The masses and decay constants are given in GeV.}
\label{inputs}
\end{table}

\begin{table}
\begin{tabular}{lccccccccc}
\hline
                     &   $m_c$  &  $A_0^{B\to K^*}$ &  $f_{K^*}$ &  $f_{K^*}^\bot(2{\rm GeV})$ & $\mu$  & $\alpha_1(K^*)$ & $\alpha_2(K^*)$ & $\alpha_1^\bot(K^*)$ & $\alpha_2^\bot(K^*)$  \\
\hline\hline
$\Delta^d_{K^*K^*}$  &  37.3\%  &       13.2\%      &    0.2\%   &            0\%              & 44.2\% &        0.1\%    &        4.6\%    &           0.1\%      &     0.3\%         \\
\hline
$\Delta^s_{K^*K^*}$  &  37.5\%  &       12.9\%      &    0.2\%   &            0\%              & 44.4\% &        0.1\%    &        4.7\%    &           0.1\%      &     0.3\%         \\
\hline\\
\end{tabular}
\begin{tabular}{lcccccccccccc}
\hline
                        &   $m_c$  &  $A_0^{B\to K^*}$  &  $f_{K^*}$ &  $f_{K^*}^\bot(2{\rm GeV})$ & $\mu$  & $\alpha_1(K^*)$ & $\alpha_2(K^*)$ & $\alpha_2^\bot(K^*)$  & $A_0^{B\to \phi}$ & $f_{\phi}^\bot(2{\rm GeV})$ & $\alpha_2(\phi)$ & $\alpha_2^\bot(\phi)$\\
\hline\hline
$\Delta^d_{\phi K^*}$   &  44.2\%  &        2.0\%       &    ---     &           ---               & 52.3\% &        ---      &        ---      &      ---              &          ---       &           0.4\%            &       0.7\%      &    0.3\%   \\
\hline
$\Delta^s_{\phi K^*}$   &  35.0\%  &        ---         &    0.1\%   &            0.7\%            & 58.2\% &        0.7\%    &        0.1\%    &     0.1\%             &          5.0\%     &           0\%              &       0\%        &    0\%   \\
\hline
$\Delta^s_{\phi \phi}$  &  44.1\%  &        ---         &    ---     &           ---               & 52.3\% &        ---      &        ---      &       ---             &          2.1\%     &           0.4\%            &       0.7\%      &    0.3\%  \\
\hline
\end{tabular}
\caption{Relative contributions from the inputs to the errors in $\Delta$ for the various decays.
}
\label{errs}
\end{table}

In a similar way, we can compute the corresponding
longitudinal $\Delta$ for the decay modes $B_{d,s}\to \phi\bar{K}^{*0}$ and $B_s\to \phi\phi$:
\eqa{
|\Delta^d_{\phi K^*}|&=&A_{K^*\phi}^{d,0} \frac{C_F \alpha_s}{4\pi N_c}C_1\,|\bar{G}_{\phi}(s_c)-\bar{G}_{\phi}(0)|=(1.02 \pm 1.11)\times 10^{-7}\ {\rm GeV} \label{delta0phiK}\\
|\Delta^s_{\phi K^*}|&=&A_{\phi K^*}^{s,0} \frac{C_F \alpha_s}{4\pi N_c}C_1\,|\bar{G}_{\phi}(s_c)-\bar{G}_{\phi}(0)|=(1.16 \pm 1.05)\times 10^{-7}\ {\rm GeV} \label{delta0sphiK}\\
|\Delta^s_{\phi\phi}|&=&A_{\phi\phi}^{s,0} \frac{C_F \alpha_s}{4\pi N_c}C_1\,|\bar{G}_{\phi}(s_c)-\bar{G}_{\phi}(0)|=(2.06 \pm 2.24)\times 10^{-7}\ {\rm GeV} \label{delta0phiphi}
}

In the following Sections we show how to apply the results of Sections
\ref{sec:pt} and \ref{sec:Delta} to the longitudinal contribution
of penguin-dominated $B\to VV$ modes. We will see that they can be
used to extract the $B_s-\bar{B_s}$ mixing angle and some
longitudinal observables like branching ratios and time-dependent
CP asymmetries within the Standard Model. In particular,
we outline three different strategies to
determine the $B_s-\bar{B_s}$ mixing angle (in the SM and beyond). Indeed,  concerning New Physics we will see that under the
assumption of no significant New Physics affecting the amplitude,
while Strategy II can detect the presence of New Physics by comparing the obtained $\phi_s$ with
$\phi_s^{SM}=2 \beta_s$, Strategy I and III can not only detect New Physics but allow also for the
extraction of $\phi_s$ even in the presence of New Physics in the mixing.

\section{FIRST STRATEGY TO EXTRACT $\phi_s$: BOUNDING $T/P$}
\label{sec:bounding}

The $b\to s$ penguin-dominated decays like $B_s\to K^{*0}
\bar{K}^{*0}$ are in principle clean modes to extract the mixing
angle $\phi_s$. In this section and those following, $\phi_s$ refers
to the same mixing angle that will be measured, for instance, in the
mixing induced CP asymmetry of $B_s \to \psi \phi$ including
possible New Physics contributions in the mixing. When focusing only
on SM we will use the notation $\phi_s=2\beta_s$.

In an expansion in powers of $\lambda_u^{(s)}/\lambda_c^{(s)}$,  the amplitude for
the decay $B_s\to K^{*0} \bar{K}^{*0}$  is given by:
\eq{\label{pr}\Amix\lg(B_s\to K^{*0} \bar{K}^{*0})\simeq \sin{\phi_s}+2\left|
\frac{\lambda_u^{(s)}}{\lambda_c^{(s)}} \right|
{\rm Re}\left( \frac{T^s_{K^*K^*}}{P^s_{K^*K^*}} \right) \sin{\gamma}\cos{\phi_s}+\cdots}
In order to determine the accuracy of this relation, we must assess
the size of the CKM-suppressed hadronic contribution $T$.
Notice that this relation is valid even in presence of New Physics in the mixing.
In the SM, one can derive from the Wolfenstein parametrisation that
eq.~(\ref{pr}) is of order $\lambda^2$ (with $\lambda=V_{us}$),
and both pieces shown on the r.h.s of eq.({\ref{pr}})
are of this same order. However, despite the smallness of the ratio
$|\lambda_u^{(s)}/\lambda_c^{(s)}|=0.044$, a significant value of
the hadronic ratio ${\rm Re}(T/P)$  could spoil the potentially safe
extraction of $\sin{\phi_s}$ (a similar issue was discussed in
ref.~\cite{bigP} for $B\to\pi\pi$). The deviation from  $\sin \phi_s$ is:
\eq{\label{deltass}\Delta S(B_s\to K^{*0} \bar{K}^{*0})\equiv 2\left|
\frac{\lambda_u^{(s)}}{\lambda_c^{(s)}} \right|
{\rm Re}\left( \frac{T^s_{K^*K^*}}{P^s_{K^*K^*}} \right) \sin{\gamma}\cos{\phi_s}}
We want to set bounds on ${\rm Re}(T/P)$, which can be related to
the inputs:
\eq{ {\rm Re}\left( \frac{T}{P} \right)={\rm Re}\left(
\frac{P+\Delta}{P} \right)=1+{\rm Re}\left( \frac{\Delta}{P}
\right)= 1+\frac{{\rm Re}(P)\,\Delta}{{\rm Re}(P)^2+{\rm Im}(P)^2}
\label{ReP/T}}
Eqs.~(\ref{eqxP}) and (\ref{eqyP}) show that the maximum of ${\rm
Re}(T/P)$ is reached for $\Adir\lg=0$ together with the positive
branch for ${\rm Re}(P)$ in Eq.~(\ref{eqxP}). The following bound is
obtained
\eq{
{\rm Re}\left( \frac{T}{P} \right)\le\,1+\left(-c_1^{(s)}+\sqrt{-(c_0^{(s)}/c_2^{(s)})^2+(1/c_2^{(s)})\,\widetilde{BR}/\Delta^2}\right)^{-1}
\label{formula}
}
where the lower bound for $BR\lg$ and the upper bound for $\Delta$
must be used. In a similar way, the minimum of ${\rm Re}(T/P)$
occurs for $\mathcal{A}_{dir}\lg=0$, for the negative branch of
Eq.~(\ref{eqxP}) for the solution of ${\rm Re}(P)$
\eq{
{\rm Re}\left( \frac{T}{P} \right)\ge\,1+\left(-c_1^{(s)}-\sqrt{-(c_0^{(s)}/c_2^{(s)})^2+(1/c_2^{(s)})\,\widetilde{BR}/\Delta^2}\right)^{-1}
\label{formula2}
}
where the lower bound for $BR\lg$ and the upper bound for $\Delta$ must be used
once again. As a conclusion, we obtain a range for ${\rm Re}(T/P)$ from
two inputs: the branching ratio $BR\lg(B_s\to K^{*0} \bar{K}^{*0})$
and $\Delta^s_{K^*K^*}$, given in Eq.(\ref{delta0sKK}).

\begin{figure}
\begin{center}
\psfrag{S}{\hspace{-1cm}$\Delta S(B_s\to K^{*0} \bar{K}^{*0})$}
\psfrag{BR}{\hspace{-1cm}$BR\lg(B_s\to K^{*0} \bar{K}^{*0})\times 10^6$}
\psfrag{10}{$10$}\psfrag{20}{$20$}\psfrag{30}{$30$}\psfrag{40}{$40$}\psfrag{50}{$50$}
\psfrag{0.03}{\hspace{0.3cm}$0.03$}\psfrag{0.035}{\hspace{0.35cm}$0.035$}\psfrag{0.04}{\hspace{0.3cm}$0.04$}
\psfrag{0.045}{\hspace{0.35cm}$0.045$}\psfrag{0.05}{\hspace{0.3cm}$0.05$}
\psfrag{0.055}{\hspace{0.35cm}$0.055$}\psfrag{0.06}{\hspace{0.3cm}$0.06$}
\includegraphics[height=7cm]{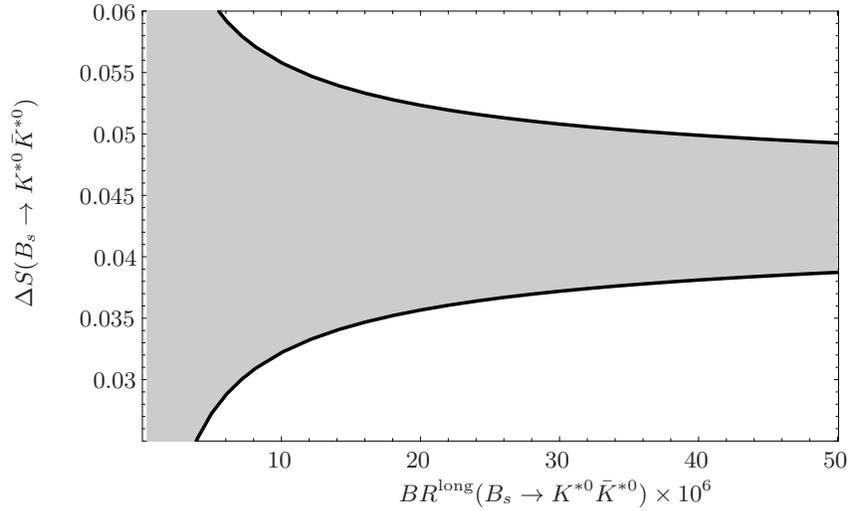}
\end{center}
\caption{Absolute bounds on $\Delta S(B_s\to K^{*0} \bar{K}^{*0})$
as a function of $BR\lg(B_s\to K^{*0} \bar{K}^{*0})$.} \label{DeltaS}
\end{figure}

Using Eq.~(\ref{deltass}), these upper and lower bounds on ${\rm
Re}(T/P)$ are converted into a bound on the pollution $\Delta
S(B_s\to K^{*0}\bar{K}^{*0})$. The latter is plotted as a function
of the longitudinal $BR\lg(B_s\to K^{*0} \bar{K}^{*0})$ in
Fig.\ref{DeltaS}.

Once a measurement of  $\Amix\lg(B_s\to K^{*0} \bar{K}^{*0})$ is available,
upper and lower bounds for
 $\phi_s$ are easily obtained.
For instance, if we take as a lower bound for the branching ratio
$BR\lg(B_s\to K^{*0} \bar{K}^{*0})\gtrsim 5\times 10^{-6}$,
Fig.~\ref{DeltaS} gives $0.03<\Delta S(B_s\to K^{*0}
\bar{K}^{*0})<0.06$.  In the case of a moderately large branching
ratio $BR\lg(B_s\to K^{*0} \bar{K}^{*0})\sim (30-40)\,\times
10^{-6}$, the bounds get sharper, with $0.04<\Delta S(B_s\to
K^{*0} \bar{K}^{*0})<0.05$ and
\eq{\big(\Amix\lg(B_s\to K^{*0} \bar{K}^{*0})-0.05\big)\ < \sin{\phi_s} <\ \big(\Amix\lg(B_s\to K^{*0} \bar{K}^{*0})-0.04\big)}

The same strategy can be applied to $B_s \to \phi K^{*0}$ and $B_s\to\phi\phi$ decays
\begin{itemize}
\item Take the experimental value for the longitudinal branching ratio $BR\lg$
  (once available), and the theoretical value for
$\Delta$ from Eqs.~(\ref{delta0sphiK}) or (\ref{delta0phiphi}).
\item Apply Eqs.~(\ref{formula}) and (\ref{formula2}) to constrain the range of ${\rm Re}(T/P)$.
\item Derive the allowed range for $\Delta S$ according to the equivalent of (\ref{deltass})
\item From the measured value of $\Amix\lg$, determine $\phi_s$ from
\eq{\big(\Amix\lg-\Delta S_{max}\big)\ < \sin{\phi_s} <\ \big(\Amix\lg-\Delta S_{min}\big)}
\end{itemize}
A weak mixing angle $\phi_s$ different from $\phi_s^{SM}$ would signal
the presence of New Physics.

Interestingly, if the longitudinal direct CP asymmetry becomes
available and happens to be inconsistent with zero, the bounds for
${\rm Re}(T/P)$ in Eq.~(\ref{formula}) and (\ref{formula2}) can be
tightened. Eq.~(\ref{ReP/T}) can be exploited to derive
expressions similar to Eq.~(\ref{formula}) and (\ref{formula2})
with a non-vanishing $\Adir\lg$, leading to stronger bounds on
${\rm Re}(T/P)$ and consequently on $\sin \phi_s$.

\section{SECOND STRATEGY :  MEASURING CP ASYMMETRIES AND BRANCHING RATIO}
\label{sec:measuring}

In this section, we show how we can extract  mixing angles and
related CKM phases in a clean way from experimental data, the
length of two sides of the unitarity triangle and the theoretical
quantity $\Delta$. The only theoretical requirement is that the
decay must allow for a safe way of computing  $\Delta$. The
approach is general in the same sense as in the previous section,
since it can be
applied to any B decay into two pseudoscalars or vectors. But it yields
different results for the four groups of decays:
\begin{enumerate}
\item $B_d$ decay through a $b\to d$ process, e.g. $B_{d}\to K^{*0} \bar{K}^{*0}$
\item $B_s$ decay through a $b\to s$ process, e.g. $B_{s}\to K^{*0} \bar{K}^{*0}$
\item $B_d$ decay through a $b\to s$ process, e.g. $B_{d}\to \phi
  \bar{K}^{*0}$ (with a subsequent decay into a CP eigenstate)
\item $B_s$ decay through a $b\to d$ process, e.g. $B_{s}\to \phi
  \bar{K}^{*0}$
(with a subsequent decay into a CP eigenstate)
\end{enumerate}
As far as weak interactions are concerned,
the difference between $B_d$ and $B_s$ decays consists in the mixing
angle, whereas $b\to d$ and $b\to s$ processes differ
through the CKM elements $\lambda_{u,c}^{(D)}$, where $D=d$ or $s$.

In the case of a $B_d$ meson decaying through a $b\to D$
process $(D=d,s)$, we can extract the angles $\alpha$ \cite{DILM}
and $\beta$ from the identities:
\eqa{
\sin^2{\alpha}&=&\frac{\widetilde{BR}}{2|\lambda_u^{(D)}|^2|\Delta|^2}\left( 1-
\sqrt{1-{(\Adir)}^2-{(\Amix)}^2} \right)\\
\sin^2{\beta}&=&\frac{\widetilde{BR}}{2|\lambda_c^{(D)}|^2|\Delta|^2}\left( 1-\sqrt{1-
{(\Adir)}^2-{(\Amix)}^2} \right)
}
In the case of a $B_s$ meson decaying through a $b\to D$ process
$(D=d,s)$, we can extract the angles $\beta_s$
\cite{quim} and $\gamma$, assuming no New Physics in the decay, from
the following expressions:
\eqa{ \label{sbetas}
\sin^2{\beta_s}&=&\frac{\widetilde{BR}}{2|\lambda_c^{(D)}|^2|\Delta|^2}\left(
1-
\sqrt{1-{(\Adir)}^2-{(\Amix)}^2} \right)\\
\sin^2{\left(\beta_s+\gamma\right)}&=&
\frac{\widetilde{BR}}{2|\lambda_u^{(D)}|^2|\Delta|^2}\left( 1-
\sqrt{1-{(\Adir)}^2-{(\Amix)}^2} \right)
}
If the obtained $\beta_s$ differs from its SM value, this would signal the presence
of New Physics.
Notice that this strategy is obtained by combining the definition of $\Delta$ with
the unitarity of the CKM matrix, so it is designed to work only in the context of
the SM. Consequently the previous expressions should be understood as a way of
testing the SM. This is an important difference with Strategies I and III where one
can obtain a value for the weak mixing phase also in the presence of New Physics in
the mixing (but not in the decay).

While the previous equations are quite general (they can be used for $B\to PP$
decays), it is understood that $BR$ and $A_{\rm dir,mix}$ refer to the
longitudinal branching ratio and longitudinal CP-asymmetries, respectively,
when they are applied to $B\to VV$ decays.

Eq.~(\ref{sbetas}) provides a new way to perform a consistency test for the SM value
of  $|\sin \beta_s |$
from the measurements of $\Amix\lg(B_s\to K^{*0} \bar{K}^{*0})$, $\Adir\lg(B_s\to K^{*0} \bar{K}^{*0})$ and $BR\lg(B_s\to K^{*0}
\bar{K}^{*0})$.
The same strategy can be applied to
$B_s\to \phi \bar{K}^{*0}$ and $B_s\to \phi\phi$ using the corresponding sum rules.
This sum rule offers several advantages : it is independent of CKM angles, and all the
hadronic input is concentrated on a single well-controlled
quantity $\Delta$.

Note that all these equations depend actually on the corresponding
branching ratio and $\mathcal{A}_{\Delta \Gamma}\lg$. The
asymmetry $\mathcal{A}_{\Delta \Gamma}\lg$ is indeed related to
the direct and mixing-induced CP-asymmetries through the equality
${(\Adir\lg)}^2+{(\Amix\lg)}^2+{(\mathcal{A}_{\Delta
\Gamma}\lg)}^2 = 1$. It was already noticed in \cite{fm} in the
context of $B_s \to K^+ K^-$ and in \cite{df} in the context of $B
\to J/\psi K^*, D^{\ast +}_s \bar D^{\ast}$ decays that it is
possible to extract $\mathcal{A}_{\Delta \Gamma}\lg$ directly from
the ``untagged'' rate:
\begin{equation}
\Gamma\lg(B_s(t)\to VV)+\Gamma\lg(\overline{B_s}(t)\to VV)
\propto R_{\rm H}e^{-\Gamma_{\rm H}^{(s)}t}+ R_{\rm
L}e^{-\Gamma_{\rm L}^{(s)}t}
\end{equation}
If the time dependence of both exponentials can be separated, one obtains
\begin{equation}
\mathcal{A}_{\Delta\Gamma}\lg(B_s\to VV)= \frac{R_{\rm H}-R_{\rm
L}}{R_{\rm H}+R_{\rm L}},
\end{equation}
The branching ratio and $\mathcal{A}_{\Delta\Gamma}\lg$ are thus
the only required observables to extract $\beta_s$ through this method, which
offers the advantage of concentrating in $\Delta$ all the hadronic input
needed to bound the tree-to-penguin ratio.

\section{THIRD STRATEGY : RELATING $B_s\to K^{*0} \bar{K}^{*0}$ AND $B_d\to
  K^{*0} \bar{K}^{*0}$}
\label{sec:flavour}

Once an angular analysis of $B_d\to K^{*0} \bar{K}^{*0}$ is
performed, it is possible to extract the CP-averaged branching
ratio corresponding to the longitudinal helicity final state.
Eqs.~(\ref{eqxP}) and (\ref{eqyP}) can be used to extract the
hadronic parameters, if one assumes that no New Physics contributes in an
appreciable way. If flavour symmetries are sufficiently accurate
for this particular process, this estimate can be converted into a
fairly precise determination of hadronic parameters for the $b\to
s$ channel $B_s\to K^{*0} \bar{K}^{*0}$. For $B_{d,s}\to KK$
modes~\cite{DMV}, we noticed that $U$-spin analysis combined with
QCD factorisation led to tight constraints on the ratio of the tree
contributions to both decay modes, as well as that for the
penguins. In this section we show how to relate $B_d\to K^{*0}
\bar{K}^{*0}$ and $B_s\to K^{*0} \bar{K}^{*0}$ decay modes
following the same approach.


We define the
parameters $\delta_{\sss K^*K^*}^P$ and $\delta_{\sss K^*K^*}^T$ as
\begin{equation}
\begin{array}{rclcrcl}
P_{K^*K^*}^{s}&=& f\,P_{K^*K^*}^{d}(1+\delta_{\sss K^*K^*}^P)\ ,&\quad &
T_{K^*K^*}^{s}&=& f\,T_{K^*K^*}^{d}(1+\delta_{\sss K^*K^*}^T)\\
\end{array}
\label{TsPs}
\end{equation}
where the factor $f$ is given by
\eq{\frac{m_{B_s}^2A_0^{B_s\to K^*}}{m_B^2A_0^{B\to K^*}}=0.88\pm 0.19}

We compute $|\delta_{\sss K^*K^*}^{P,T}|$ using QCDF. These parameters are affected by the model dependent treatment of annihilation
and spectator-scattering contributions, so
the results should be considered
 as an estimate. A significant part of long-distance dynamics is common
to both decays, and we find the following upper bounds
\begin{equation}
|\delta_{\sss K^*K^*}^P|\le 0.12\ , \qquad |\delta_{\sss K^*K^*}^T|\le 0.15
\end{equation}
where the largest contribution comes from the lower value of $\lambda_B$.

We could in principle apply the same strategy to $B_{d,s} \to \phi
K^{*0}$, but the corresponding $\delta$'s are much larger. Indeed,
the computation leads to corrections up to $\delta_{\sss \phi
K^*}\sim 50\%$. This shows that $U$-spin symmetry cannot be
expected to hold at a high accuracy for any pair of
flavour-related processes. $K^{(*)}K^{(*)}$ offer a much more
interesting potential than other final states such as $\phi
K^{*0}$. Moreover, we cannot perform a similar analysis for
$\phi\phi$ since $B_d\to\phi\phi$ is a pure weak-annihilation
process, contrary to $B_s\to\phi\phi$ mediated through penguins.
Therefore we focus on the  precise $B_s \to K^{*0} {\bar
K^{*0}}$ modes in the remaining part of this section. Notice that
the large hadronic uncertainties affecting $B_s \to \phi\phi$ and
$B_s \to \phi K^{*0}$ have no impact when we use these modes in
the strategies described in Sec. V and VI, since we exploited a
quantity $\Delta$ where they cancel out.

%
%

\begin{figure}
\psfrag{BRd}{\hspace{-1.5cm}$BR\lg(B_d\to K^{*0} \bar{K}^{*0})\times 10^6$}
\psfrag{BRs}{\hspace{-1.5cm}$BR\lg(B_s\to K^{*0} \bar{K}^{*0})\times 10^6$}
\includegraphics[height=7cm]{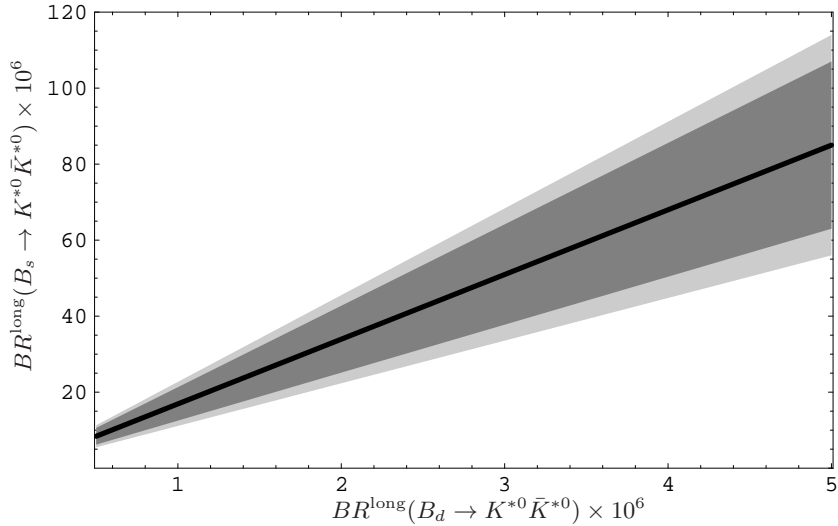}
\caption{Longitudinal branching ratio for $B_s\to K^{*0} \bar{K}^{*0}$
in terms of the longitudinal $B_d\to K^{*0} \bar{K}^{*0}$
branching ratio. The light-shaded area corresponds to the uncertainty on
the ratio of form factors $f$, whereas the dark-shaded area comes from
varying the various hadronic inputs.} \label{BstoKK}
\end{figure}

Once the hadronic parameters $P_{K^*K^*}^{s}$ and $T_{K^*K^*}^{s}$ have
been obtained from Eq.(\ref{TsPs}), one can
give predictions for the $B_s\to K^{*0} \bar{K}^{*0}$ observables.
Note that the branching ratio $BR\lg(B_d\to K^{*0} \bar{K}^{*0})$
is an experimental input in this analysis, and this piece of information is not  available yet.
The result for the branching ratio of $B_s\to K^{*0} \bar{K}^{*0}$ is
given in terms of the $B_d\to K^{*0} \bar{K}^{*0}$ branching ratio
in Fig.\ref{BstoKK}. Once the branching ratio of $B_d\to K^{*0} \bar{K}^{*0}$
is measured one can use this plot to find the SM prediction
for $BR\lg(B_s\to K^{*0} \bar{K}^{*0})$.

The ratio of branching ratios $BR\lg(B_s\to K^{*0}\bar{K}^{*0})/BR\lg(B_d\to
K^{*0}\bar{K}^{*0})$ and the asymmetries turn out to be quite insensitive to
the exact value of  $BR\lg(B_d\to K^{*0}\bar{K}^{*0})$ as long as
$BR\lg(B_d\to K^{*0}\bar{K}^{*0})\gtrsim 5 \times 10^{-7}$. The
numerical values are summarised in Table~\ref{Tableresults}.

\begin{table}
\begin{tabular}{|c|}
\hline
$\begin{array}{c}
\\
\quad\displaystyle
\left(\frac{BR\lg(B_s\to K^{*0}\bar{K}^{*0})}{BR\lg(B_d\to K^{*0}\bar{
K}^{*0})}\right)_{\sss SM}=17\pm 6 \quad\\
\\
\end{array}$\\
\hline
$\begin{array}{c}
\\
\Adir\lg(B_s\to K^{*0}\bar{K}^{*0})_{\sss SM}=0.000\pm 0.014\\
\\
\end{array}$\\
\hline
$\begin{array}{c}
\\
\Amix\lg(B_s\to K^{*0}\bar{K}^{*0})_{\sss SM}=0.004\pm 0.018\\
\\
\end{array}$\\
\hline
\end{tabular}
\caption{Results for the longitudinal observables related to $B_s\to
  K^{*0}\bar{K}^{*0}$ according to Sec.~\ref{sec:flavour}.
These are predictions for the SM contributions under the standard
assumption of no New Physics in $b\to d$ transition. We used
$\phi_s^{SM}=2 \beta_s=-2^\circ$ for $\Amix\lg$, and
we assumed $BR\lg(B_d\to K^{*0}\bar{K}^{*0})\gtrsim 5 \times
10^{-7}$. The quoted uncertainty includes the errors associated to all input parameters including
 the variation of $\gamma$ inside the range $56^\circ \leq \gamma\ \leq 68^\circ$\cite{charles}.}
\label{Tableresults}
\end{table}

Under the standard assumption that  New Physics contribution to $b \to d$ penguins is negligible, and
since the experimental
input comes entirely from $B_d\to K^{*0}\bar{K}^{*0}$ (a $b \to d$ penguin), the results given in Table \ref{Tableresults} are
SM predictions.
In  presence of New Physics in $b\to s$ penguins the full prediction can be
obtained by adding to
the SM piece extra
contributions to the
amplitude and weak mixing angle as explained in \cite{LMV,extra}.

\begin{figure}
\psfrag{phis}{\hspace{-0.5cm}$\phi_s$ (Degree)}
\psfrag{Amix}{\hspace{-1cm}$\Amix\lg(B_s\to K^{*0}\bar{K}^{*0})$}
\includegraphics[height=7cm]{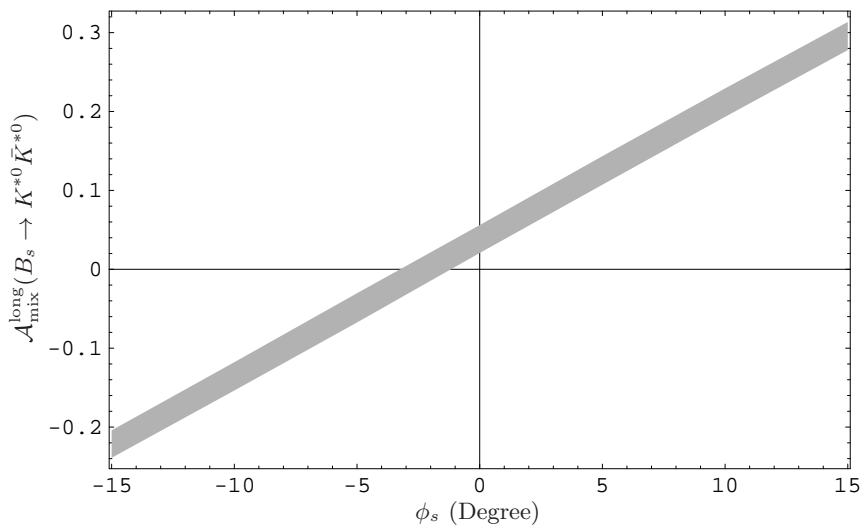}
\caption{Relation between $\Amix\lg(B_s\to K^{*0}\bar{K}^{*0})$
and the $B_s-\bar{B}_s$ mixing angle $\phi_s$. We assumed
$BR\lg(B_d\to K^{*0}\bar{K}^{*0})\gtrsim 5 \times 10^{-7}$ and
$\gamma=62^\circ$. A measurement of this asymmetry leads to a
prediction for $\phi_s$, which includes hadronic pollution and
SU(3) breaking effects, according to Sec.~\ref{sec:flavour}.}
\label{Amix-phis}
\end{figure}

One may also use this as a strategy to extract the mixing angle
$\phi_s$. If one assumes no New Physics in the decay $B_s\to
K^{*0}\bar{K}^{*0}$, this method relates directly $\Amix\lg(B_s\to
K^{*0}\bar{K}^{*0})$ and $\phi_s$. Fig.\ref{Amix-phis} shows
$\Amix\lg(B_s\to K^{*0}\bar{K}^{*0})$ vs. $\phi_s$. Once this
asymmetry is measured, this plot can be used as a way to extract
$\phi_s$, and this result can be compared to the one found in tree
decays such as $B\to DK$. A disagreement would point out New
Physics. Moreover, it is possible to distinguish whether New
Physics affects the decay or the mixing itself, by confronting
$BR\lg(B_s\to K^{*0}\bar{K}^{*0})$ and $\Adir\lg(B_s\to
K^{*0}\bar{K}^{*0})$ with the SM predictions given in
Table~\ref{Tableresults}. If the predictions for the branching
ratio and the direct CP asymmetry agree with experiment, but the
$\phi_s$ extracted from $\Amix\lg(B_s\to K^{*0}\bar{K}^{*0})$
differs from $\phi_s^{SM}$, this will be a clear indication of New
Physics in $B_s-\bar{B}_s$ mixing. An interesting comparison will
be allowed between the value for $\phi_s$ obtained here and the
measurement of $\phi_s$ from the mixing induced CP-asymmetry of
$B_s\to D K$ decay \cite{bdk}.

\section{DISCUSSION} \label{sec:discussion}

The increasing list of measured non-leptonic two-body $B_d$- and $B_s$-decays
provides many tests of the CKM mechanism of CP violation in the
Standard Model. Of particular interest is the determination of angles through
time-dependent CP-asymmetries. For instance $\phi_s$, related to
$B_s-\bar{B}_s$ mixing, should be constrained : it is tiny in the Standard
Model, and can be measured through many penguin-dominated decays. However, for
such determination to be valid, one must assess the size of the various
hadronic quantities involved as precisely as possible.

\begin{table}
\begin{tabular}{|c||c|c|c|}
\hline
& Strategy 1 & Strategy 2 & Strategy 3\\
\hline\hline
\hspace{0.5cm}Inputs\hspace{0.5cm} &
$\begin{array}{c}
\\
BR\lg(B_s\to K^{*0} \bar{K}^{*0})\\
\Amix\lg(B_s\to K^{*0} \bar{K}^{*0})\\
\Delta^s_{K^*K^*},\ \gamma\\
\\
\end{array}$&
$\begin{array}{c}
BR\lg(B_s\to K^{*0} \bar{K}^{*0})\\
\Adir\lg(B_s\to K^{*0} \bar{K}^{*0})\\
\Amix\lg(B_s\to K^{*0} \bar{K}^{*0})\\
\Delta^s_{K^*K^*}
\end{array}$&
$\begin{array}{c}
BR\lg(B_d\to K^{*0} \bar{K}^{*0})\\
\Amix\lg(B_s\to K^{*0} \bar{K}^{*0})\\
\Delta^d_{K^*K^*},\ \delta_T,\ \delta_P,\ \gamma
\end{array}$\\
\hline
\hspace{0.5cm}Outputs\hspace{0.5cm} &
$\phi_s$&
$|\sin{\beta_s}|$, $\gamma$&
$\begin{array}{c}
\\
BR\lg(B_s\to K^{*0} \bar{K}^{*0})_{\sss SM}\\
\Adir\lg(B_s\to K^{*0} \bar{K}^{*0})_{\sss SM}\\
\phi_s\\
\\
\end{array}$\\
\hline\hline
\hspace{0.5cm}Advantages\hspace{0.5cm} &
\begin{minipage}[c]{3cm}
\vspace{0.2cm}
Applies also to $B_s\to\phi K^{*0}$ and $B_s\to\phi \phi$
\vspace{0.2cm}
\end{minipage}&
\begin{minipage}[c]{3cm}
\vspace{0.2cm}
Applies also to $B_s\to\phi K^{*0}$ and $B_s\to\phi \phi$
\vspace{0.2cm}
\end{minipage}&
\begin{minipage}[c]{3cm}
\vspace{0.2cm}
It can be easily generalized to include New Physics in the
decay and mixing.
\vspace{0.2cm}
\end{minipage}\\
\hline
\hspace{0.5cm}Limitations\hspace{0.5cm} &
\begin{minipage}[c]{3cm}
\vspace{0.2cm}
It assumes no New Physics in $b \to s$ decay.
\vspace{0.2cm}
\end{minipage}&
\begin{minipage}[c]{3cm}
\vspace{0.2cm}
It assumes no New Physics in $b \to s$ decay.
\vspace{0.2cm}
\end{minipage}&
\begin{minipage}[c]{3cm}
\vspace{0.2cm}
Does not apply to $B_s\to\phi K^{*0}$ or $B_s\to\phi \phi$
because $\delta_{T,P}$ are big.
\vspace{0.2cm}
\end{minipage}\\
\hline
\end{tabular}
\caption{Comparison between the three strategies for $B_s\to K^{*0} \bar{K}^{*0}$.}
\label{comparative}
\end{table}

In the present note, we have applied ideas presented in ref.~\cite{DMV} for
$B_{d,s}\to KK$ to vector-vector modes mediated through penguins : $B_{d,s}\to
K^{*0} \bar{K}^{*0}$, $\phi\phi$ and $\phi K^{*0}$ (with the condition that
$K^{*0}$ decays into a definite-CP eigenstate). In order to combine flavour
symmetries with QCD factorisation, we have restricted our analysis to
longitudinal observables, which are under better theoretical control. These
observables have been related to the angular analysis performed experimentally
in Sec.~\ref{sec:longobs}. Penguin-mediated modes offer the very
interesting feature that the difference between tree and penguin contributions
$\Delta=T-P$ should be dominated by short-distance physics. It can be computed fairly accurately
using QCD factorisation, and it can be used to determine
tree and penguin contributions from observables
as explained in Secs.~\ref{sec:pt} and \ref{sec:Delta}. This theoretical piece
of information is used to relate CP-asymmetries of $B_{d,s}\to
K^{*0} \bar{K}^{*0}$, $\phi\phi$ and $\phi K^{*0}$ to CKM angles according to
different strategies. For illustration, we have focused on $B_{d,s}\to
K^{*0} \bar{K}^{*0}$, where all three strategies apply.

In Sec.~\ref{sec:bounding}, we have proposed to use $\Delta=T-P$
to put stringent bounds on the pollution due to hadronic uncertainties. Indeed, even though the
ratio $|\lambda_u^{(s)}/\lambda_c^{(s)}|=0.044$ is small, a large
value of the hadronic quantity ${\rm Re}(T/P)$ could spoil the
naively safe extraction of $\sin{\phi_s}$ from the mixed asymmetry
of $B_s\to K^{*0}\bar{K}^{*0}$. This strategy to control the
pollution can be applied to all penguin-mediated processes of
interest here.

In Sec.~\ref{sec:measuring}, we have suggested a second approach, using
$\Amix\lg(B_s\to K^{*0} \bar{K}^{*0})$,
$\Adir\lg(B_s\to K^{*0} \bar{K}^{*0})$ and
$BR\lg(B_s\to K^{*0} \bar{K}^{*0})$ to extract $|{\rm sin} \beta_s|$. In
principle,
one can also use an alternative set of experimental quantities : the
branching ratio together with a direct measurement of
the longitudinal untagged rate.
The sum rule needed for the $B_s\to K^{*0}
\bar{K}^{*0}$ is independent of the CKM angle $\gamma$ and the
input on hadronic dynamics is limited to a single well-controlled
quantity: $\Delta^{s}_{K^*K^*}$.
This strategy can also be applied to extract
$\beta_s$ from $B_s\to \phi
\bar{K}^{*0}$ and $B_s \to \phi\phi$ using the corresponding sum rule.

In Sec.~\ref{sec:flavour}, we proposed a last method to
determine  $\phi_s$, by
relying on the prediction of the mixing induced CP-asymmetry
$\Amix\lg(B_s\to K^{*0} \bar{K}^{*0})$ as a function of the
$BR\lg(B_d \to K^{*0} \bar{K}^{*0})$ and the theoretical input
$\Delta^d_{K^*K^*}$. In this strategy, tree pollution is
controlled using the hadronic information from flavour symmetry
and QCD factorisation. The outcome of our analysis is presented in
Fig.~\ref{Amix-phis}.
This strategy requires data on
$B_d \to K^{*0} \bar{K}^{*0}$ and on the mixing-induced CP-asymmetry
$\Amix\lg(B_s\to K^{*0}
\bar{K}^{*0})$. The input from $B_s$ decay is therefore minimal :
 $\Amix\lg(B_s\to K^{*0} \bar{K}^{*0})$, while all
other inputs can be obtained from $B$-factories.

A comparison among the three different strategies discussed in this paper is given
in Table \ref{comparative}, where the needed inputs are enumerated
as well as the predicted observables and the range of
validity.

If both hadronic machines and super-$B$ factories \cite{Bona:2007qt} running at $\Upsilon(5S)$
provide enough information on $B_s$-decays, it will be interesting to compare the
determination from $\phi_s$ following those methods, which rely on
penguin-mediated decays, with the value obtained from tree
processes like $B_s \to DK$ \cite{bdk}. Differences between the
values obtained through these two procedures would provide a clear
hint of physics beyond the Standard Model. In such a situation,
the different methods presented in this letter would yield very
useful cross-checks for the penguin-dominated vector modes.

\begin{acknowledgments}

We thank B.~Adeva, G.~Punzi and S.~T'Jampens for useful discussions.
This work was supported in part by the EU Contract No. MRTN-CT-2006-035482, \lq\lq FLAVIAnet''.
J.M acknowledges support from RyC
program (FPA2002-00748 and PNL2005-41).

\end{acknowledgments}

\end{document}